\documentclass{article}

\usepackage{arxiv}

\usepackage[utf8]{inputenc} % allow utf-8 input
\usepackage[T1]{fontenc}    % use 8-bit T1 fonts
\usepackage{hyperref}       % hyperlinks
\usepackage{url}            % simple URL typesetting
\usepackage{booktabs}       % professional-quality tables
\usepackage{amsfonts}       % blackboard math symbols
\usepackage{nicefrac}       % compact symbols for 1/2, etc.
\usepackage{microtype}      % microtypography
\usepackage{lipsum}
\usepackage{graphicx}
\graphicspath{ {./images/} }

\title{Simplifying Sparse Expert Recommendation by Revisiting Graph Diffusion}

\author{
  Vaibhav Krishna, Nino Antulov-Fantulin  \\
  ETH Zürich \\
  email: \texttt{vaibhavkrishna@ethz.ch},\\   \texttt{nino.antulov@gess.ethz.ch}
}

\begin{document}
\maketitle
\begin{abstract}
Community Question Answering (CQA) websites have become valuable knowledge repositories where individuals exchange information by asking and answering questions. With an ever-increasing number of questions and high migration of users in and out of communities, a key challenge is to design effective strategies for recommending experts for new questions. 
In this paper, we propose a simple graph-diffusion expert recommendation model for CQA, that can outperform state-of-the art deep learning representatives and collaborative models.  Our proposed method learns users' expertise in the context of both semantic and temporal information to capture their changing interest and activity levels with time. 
Experiments on five real-world datasets from the Stack Exchange network demonstrate that our approach outperforms competitive baseline methods. Further, experiments on cold-start users (users with a limited historical record) show our model achieves an average of $\approx 30\%$ performance gain compared to the best baseline method. 
\end{abstract}


% keywords can be removed
%\keywords{First keyword \and Second keyword \and More}

\section{Introduction}
 Community-based question answering sites are becoming increasingly important for sharing and obtaining knowledge \cite{srba2016comprehensive}. Popular CQA websites like Quora, Stack Overflow, and Stack Exchange sites may have tens of thousands of questions posted every day. This led to a severe gap between the posted questions and potential experts who can provide answers, resulting in as much as 30\% of the questions remaining unanswered \cite{srba2016stack}. 

To overcome this gap existing studies have proposed methods to recommend experts for newly posted questions \cite{li2011question}. This improves the chances for the questions to receive a high-quality answers since the recommended users are able to immediately spot the questions of their expertise. The success of expert recommendation can potentially increase the participation rates of users and foster stronger communities in CQA. Given their advantages such systems have attracted lot of attention in the information retrieval community\cite{neshati2017dynamicity}.

One of the solution to address the problem of expert recommendation is addressed by feature engineering-based approaches\cite{zhou2012classification}. These algorithms extract features from users, questions and their relations and feed them to models like SVM and linear regression \cite{chang2013routing} to rank the expertise of users. However, these methods are time consuming and suffer from selection bias as hand-crafted features are needed. Another approach is network based which utilize link analysis technique on a user-user network form from asking-answering relationship to evaluate the authority of users and rank them \cite{zhang2007expertise}. However, both the above approaches ignores the topic expertise of users and recommend them based on general expertise.

To overcome this gap text-based approaches are proposed that model users' topical knowledge based on topics extracting from previous Q\&A posts using language and topic models \cite{zhou2012topic,riahi2012finding}. In addition, certain hybrid approaches have been proposed that integrate network approaches and features extracted from text \cite{li2015hybrid,yang2013cqarank}. This makes these approaches computationally expensive. Another set of studies that have shown promising results in modeling experts in CQAs are based on matrix factorization techniques which are known for their advantages of flexibility and scalability in the recommendation domain \cite{zhao2014expert}.

However, despite the active research in CQA, the expert recommendation remains a challenging task due to reasons like sparsity of historical question and answer records, low participation rates of users, migration of users in and out of the community, and lack of personalised recommendation, etc \cite{wang2018survey}. While existing approaches address some of these issues, these approaches face challenges in three aspects. First, a lot of existing approaches use all the content of past questions and answers posted by the users to estimate their topical expertise for personalised recommendations. This makes these approaches computationally expensive. Second, most of the existing approaches consider a static view of these platforms. However, CQA platforms are highly dynamic with high migration of users in and out of the platform, as well as changing interests and activity patterns of existing users \cite{pal2012evolution}. Third, of the answers that are posted each month on these platforms, around one-third answers are posted by new users with limited past answering records. For the existing users, a lot of users have a low participation rate leading to high sparsity \cite{le2016retrieving}. This results in a key challenge - that of cold-start users.

In order to address the challenges above, we propose a novel \underline{t}emporal weighted \underline{b}ipartite \underline{g}raph model for \underline{e}xpert \underline{r}ecommendation (t-BGER). Given the importance of question tags in previous studies, we focus on learning the user topic expertise based on tags. Past studies have highlighted the advantages of using tags to find topic expertise \cite{yang2014tag} that outperformed other text-based topic modeling approaches \cite{yang2013cqarank} and was several magnitudes faster. Thus, focusing on tags to identify topical expertise reduces the computational complexity of our approach, making it easier to implement for periodic updates. Further, by incorporating the temporal information of the posts our approach indirectly models users evolving interests and activity patterns. Finally, we use a bipartite graph approach which is more accurate and faster compared to matrix decomposition techniques as well as achieves better performance when working with highly sparse data \cite{zhou2010solving}. We evaluate our proposed approach on five different CQA datasets from the Stack Exchange network. The results demonstrate that our approach outperforms competing baseline approaches on multiple metrics.

The rest of the paper is organised as follows. Section 2 presents our proposed approach. Next in Section 3 we present the data and experimental results. Finally, we conclude in Section 4.

 \section{Our Method}
In this section, we will introduce our approach referred to as t-BGER. We first introduce some concepts.

\begin{figure}[h]
  \centering
  \includegraphics[scale=0.6]{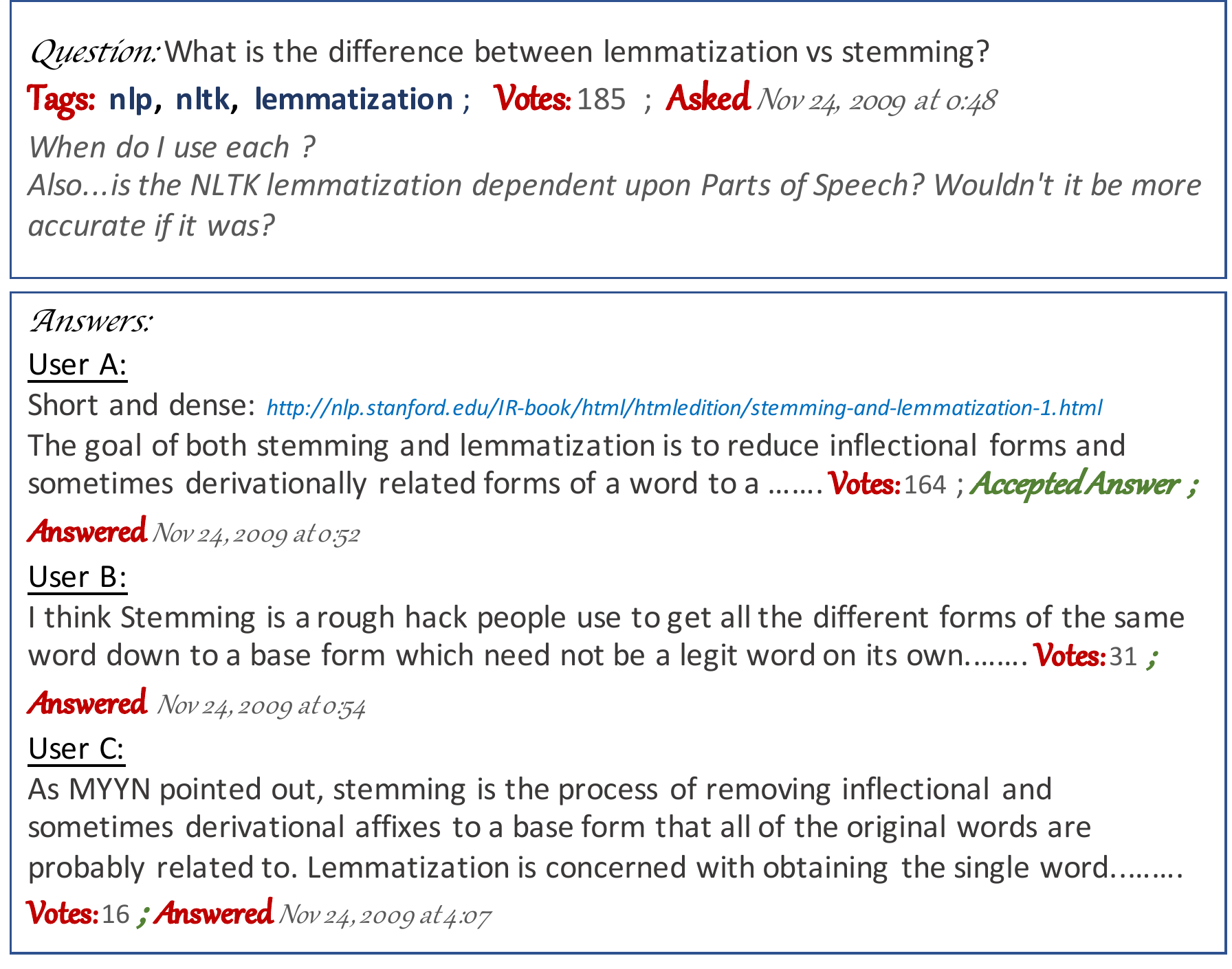}
  \caption{Sample timestamped Q\&A post with tags, votes and accepted answer.}
  \label{fig1}
\end{figure}

\textbf{User}: We use \textit{users} to refer to the seekers and answerers in CQA. Figure \ref{fig1} shows a snapshot of typical Q\&A posts with corresponding tags, votes, answerers, and accepted answer. \textbf{Tags}: The seeker can assign a maximum of five tags to the question that summarizes the question focus. \textbf{Score}: The users in CQA can vote-up or vote-down the question and answer posts. The final score equals the number of positive votes minus the number of negative votes. \textbf{Accepted answer}: The seeker can vote one of the answers as the accepted answer. The ground truth for a question is the answerer who has provided the accepted answer.

We first build a user-tag activity matrix based on the past answering activity of the users. We use tags on the questions to represent the topics as they are seen as more informative and representative in describing users' topic expertise. Next, to take into account the dynamic aspect of the CQA environment, we temporally model the past answering activities of the users. The adjacency matrix thus obtained is used to define a temporal weighted bipartite graph. Finally, potential experts' rank list is prepared using recommendations based on the bipartite graph model - a two-step probabilistic resource diffusion model based on the random walk \cite{zhou2007bipartite}.

\subsection{User-Tag Activity Matrix}
To build a user-tag activity matrix $\mathbf{S} \in\mathbb{R}^{M \times N}$, where $M$ is the number of users ($u_i$) and $N$ is the number of tags ($\nu_\tau$), we first collect all the positively scored answers by each user and the respective question along with the associated tags. Next, users are assigned an activity score on the tag of the question which is equal to $1/\verb|#|tags$. For example, in the sample question in figure \ref{fig1}, user-A gets an activity score of $1/3$ on each of the three tags. Finally, each entry of the user-tag activity matrix $s_{i \tau}$ can be obtained as the sum of all the activity scores from all the questions on the tag $\tau$ by the user $i$.

\subsection{Temporal Modelling}
To account for the dynamic behavior of users, changing interests and activity, we integrate temporal discounting in the user-tag activity matrix $\mathbf{S}$. Temporal discounting is a well-studied phenomenon in economy and psychology where people tend to discount the delayed rewards and give more weightage to near future rewards \cite{green1994temporal}. Similarly, for events in the past, people give more value to recent events. Likewise in a dynamic environment where users' activity patterns change with time, the system should give more value to recent activities. In this way, we can indirectly model and use the availability and interest of users. 

Given a CQA site where first post occurred at time $t_1=1$ and a new question $\hat{q}$ is posted at time $t_q=\hat{t}$. To identify potential answerers for the question $\hat{q}$ all the previous positively scored answers within the period $[t_1,t_q]$ by the users are considered. Next, the time period $[t_1,t_q]$ is divided into specific time windows such as day, week, or month. The answers are grouped according to their corresponding time window $\delta$ determined by the date of the answers. Next, user tag activity matrix $\mathbf{S}_\delta$ is defined for each time window, where an entry $s_{i \tau}^{\delta}$ corresponds to the user $u_i$ activity on tag $\nu_\tau$ for the time window $[t_{\delta-1},t_{\delta}]$.

Next the activity for the time window is temporally discounted using a temporal kernel (hyperbolic discounting function) of the form:

\begin{equation}
\label{eq_hyper_disc}
    \Phi(\delta)=\frac{1}{1+\delta},
\end{equation} 
where $\delta$ is the number of time windows that passed from the time of interest until the time $t_q$. For example, taking time-window as one month, for answers posted in the previous month of $t_q$, $\delta = 1$, while for answers posted ten months before $t_q$, $\delta = 10$. 

Finally, we define a temporal user-tag activity matrix $\mathbf{S}(\hat{t})$, where an entry $s_{i \tau}^{\hat{t}}$ corresponds to the user $u_i$ temporal-discounted activity on tag $\nu_\tau$.

\begin{equation}
\mathbf{S}({\hat{t}}) = \sum_{\delta=1}^{\Delta}\Phi(\delta) \mathbf{S}_{\delta},
\label{temporal_activity_Netscr}
\end{equation} 
where, ${\Delta}$ is the total number of time windows.

\subsection{Temporal Weighted Bipartite Graph}
\begin{figure}[h]
  \centering
  \includegraphics[scale=0.3]{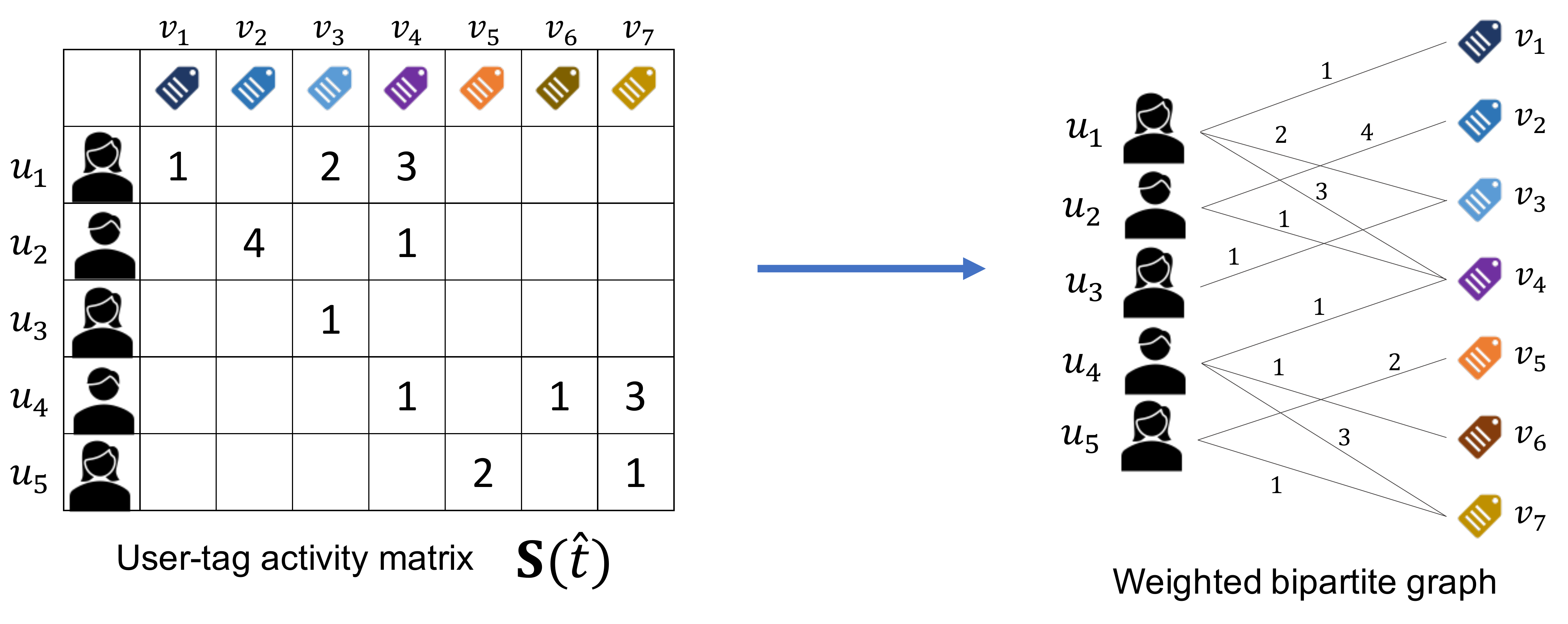}
  \caption{Sample weighted bipartite graph}
  \label{fig2}
\end{figure}

Next, using the temporal user-tag activity matrix $\mathbf{S}(\hat{t})$ we establish the temporal weighted bipartite graph $\mathcal{G}=(\mathcal{M},\mathcal{N},\mathcal{E},\mathcal{W})$, where the edge set $\mathcal{E}$ contains undirected edges and the weight set $\mathcal{W}$ contains  edge weights (Fig \ref{fig2}). The nodes in users $\mathcal{M}$ and tags $\mathcal{N}$ are denoted by $u_i$ that range from $1$ to $m$ and $\nu_\tau$ that range from $1$ to $n$ respectively. The recommender system can be describe by an $m \times n$ adjacent matrix $\{a_{i\tau}\}$, where
%We assume that a certain amount of resource (i.e. recommendation power) is associated with each user $f({u_i}) \geq 0$. 
%In the simplest case, the initial resource vector $\mathbf{f_i}$ can be set as 
%The resource located on the $u_i$ $\in$ $M$ node  is $f{u_i} \geq 0$, or on tag $\nu_\tau$ is $f(\nu_\tau) \geq 0$. 

%f_{i,\tau} =
\begin{equation}
a_{i \tau} = \begin{cases}
1,  \hspace{1cm}  u_i\nu_\tau \in \mathcal{E} \\
0,  \hspace{1cm}   otherwise
\end{cases}
\end{equation} 

That is to say, if the user $u_i$ has answered a question with tag $\nu_\tau$, then its initial resource is unit, otherwise it is zero. 

To integrate the temporal aspect, we define temporal weighted bipartite graph where

% F : matrix dimension users x tags$
% F_{i,}: vector 1 x tags
% F_{,\tau}: vector users x 1

%f_{i,\tau} = 
\begin{equation}
a_{i \tau} = \begin{cases}
s_{i \tau}^{\hat{t}},  \hspace{0.5cm}  u_i\nu_\tau \in \mathcal{E} \\
0,  \hspace{1cm}   otherwise
\end{cases}
\end{equation}

\subsection{Resource Allocation and Recommendation}
\textbf{Network-based diffusion process:}
%Once we define the temporal weighted bipartite graph with initial resources, we next perform resource allocation 
First, we define standard network-based diffusion process \cite{ou2007power}. This is done in two steps. In the first step, all the resource in $\mathcal{M}$ flows to $\mathcal{N}$, and thus the resource allocated on the ${\nu}_\tau$ $\in$ $\mathcal{N}$  reads

\begin{equation}
f(\nu_\tau) = \sum_{i=1}^{m}\frac{a_{i\tau}f(u_i)}{k(u_i)}
\end{equation} 
where $k(u_i)$ is the degree of $u_i$, $f(u_i)$ is the initial resource on node $u_i$.

In the next step, all the resource flows back to nodes in $\mathcal{M}$, and the final resource located on $u_i$ reads

\begin{equation}
f'(u_i) = 
\sum_{\tau=1}^{n}\frac{a_{i\tau}f(\nu_\tau)}{k(\nu_\tau)}=
\sum_{\tau=1}^{n}\frac{a_{i\tau}}{k(\nu_\tau)} \sum_{\beta=1}^{m}\frac{a_{\beta\tau}f(u_\beta)}{k(u_\beta)}
\end{equation} 
where $k(\nu_\tau)$ is the degree of $\nu_\tau$.

%This can be rewritten as 
%\begin{equation}
%f'(u_i) = \sum_{\beta=1}^{m}{w_{i\beta}f(u_\beta)}
%\end{equation} 
%where 
%\begin{equation}
%w_{i\beta} =  \frac{1}{k(u_\beta)} %\sum_{\tau=1}^{n}\frac{a_{i\tau}a_{\beta\tau}}{k(\nu_\tau)}
%\end{equation}
%which sums the contribution from all two-step paths between $u_i$ and $u_\beta$.

\textbf{Recommendation process}: 
The network-based diffusion process $f^{'}=Wf$, can be rewritten using a linear operator $W$, where:
\begin{equation}
w_{i\beta} =  \frac{1}{k(u_\beta)} \sum_{\tau=1}^{n}\frac{a_{i\tau}a_{\beta\tau}}{k(\nu_\tau)}
\end{equation}
which sums the contribution from all two-step paths between $u_i$ and $u_\beta$.
The weight $w_{i\beta}$ represents the proportion of the resource user $u_\beta$ would like to distribute to user $u_i$. In other words, the weight $w_{i\beta}$ contributes to the strength of recommending the user $u_i$ to a question with tag $\nu_\tau$ provided the user $u_\beta$ has answered questions with tag $\nu_\tau$. After resource-allocation process, the user-tag expertise matrix can be written as:

%the final resource vector is 

%\begin{equation}
%\mathbf{f'_i} = \mathbf{W}\mathbf{f_i}
%\label{eq9}
%\end{equation} 
%where $\mathbf{W} \in\mathbb{R}^{MxM}$ is the matrix with weights $w_{i\beta}$.

%Finally, using eq-\ref{eq9} the user-tag expertise matrix can be obtained as

\begin{equation}
\mathbf{\hat{S}({\hat{t}})} = \mathbf{W} \mathbf{{S}({\hat{t}})}
\end{equation} 

where: $\mathbf{W} \in\mathbb{R}^{M \times M}$ is the matrix with weights $w_{i\beta}$.

For a newly posted question $\hat{q}$, the tags of the question are used to compute the final score of each user using user-tag expertise matrix $\mathbf{\hat{S}({\hat{t}})}$ and finally to obtain the user rank list for recommendation.

\section{Experiments}

\subsection{Data Set}
We use five real-world CQA datasets from Stack Exchange network to evaluate the performance of proposed model. All the datasets are publically available on the archive\footnote{\url{ https://archive.org/download/stackexchange}}. The details of the datasets are presented in Table \ref{table1}. The datasets includes all the information on timestamped questions and the corresponding answer records. The questions also include the annotated tags and the answerer who provided the accepted answer. For evaluating the our approach we split the datasets in the ratio of 70:10:20 for training, validation and testing set based on the timestamped of the questions. Thus, the latest 20\% dataset in the order of question raising time is reserved as testing set. The answerer who provided the accepted answer for each question will be used as the ground truth for evaluating our approach. Following the settings in previous studies \cite{li2019personalized,zhang2020temporal}, we filtered the users who provided less than five answers out of the training set.

\begin{table}[!ht]
\caption{Summary statistics of datasets. \label{table1}}
\centering
\begin{tabular}{|l|l|l|l|l|}
\hline
CQA & Questions & Answerers & Tags & Time range\\ \hline

philosophy  & 4295 & 658  & 383 & 2011.04-2019.05  \\ \hline
  
history & 4807 & 473  & 678 & 2011.05-2019.05  \\ \hline

ebooks & 368 & 74  & 135 & 2013.12-2019.05  \\ \hline
 
3dprinting & 963 & 223  & 383 & 2016.01-2019.05  \\ \hline

%ai & 1130 & 163  & 420 & 2016.08-2019.06  \\ \hline
 
\end{tabular}
\end{table}

\begin{table*}
  \caption{Comparison of expert recommendation for different methods}
  \label{perform}
  \begin{tabular}{|l|ccl|ccl|ccl|ccl|}
    \toprule
    & \multicolumn{3}{|l|}{\bf philosophy} &
\multicolumn{3}{|l|}{\bf history} &
\multicolumn{3}{|l|}{\bf ebooks} &
\multicolumn{3}{|l|}{\bf 3dprinting} \\ 

    \midrule
    
   & MRR & P@1 & P@3 & MRR & P@1 & P@3 & MRR & P@1 & P@3 & MRR & P@1 & P@3 \\
   %& MRR & P@1 & P@3 \\
    \midrule
\bf Score &0.027 & 0.005 & 0.013 & 0.028 & 0.005 & 0.015 & 0.104 & 0.022 & 0.065 & 0.089 & 0.024 & 0.062 \\
%& 0.080 & 0.019 & 0.056 \\

\bf NeRank &0.119 & 0.055 & 0.116 & 0.124 & 0.044 & 0.130 & 0.254 & 0.160 & 0.266 & 0.135 & 0.049 & 0.130 \\ %& 0.144 & 0.061 & 0.144 \\

\bf TAG-MF & 0.156 & 0.082 & 0.146 & 0.123 & 0.034 & 0.126 & 0.430 & 0.202 & 0.566 & 0.409 & 0.259 \\ %& 0.469 & 0.357 & \bf 0.223 & 0.429 \\

\bf TCQR &0.248 & 0.135 & 0.289 & 0.206 & 0.093 & 0.224 & 0.359 & 0.190 & 0.456 & 0.461 & 0.318 & 0.528 \\ %& 0.253 & 0.137 & 0.282 \\

\bf t-TAG-MF & 0.413 & 0.297 & 0.443 & 0.201 & 0.072 & 0.199 & 0.633 & 0.338 & 0.945 & 0.631 & 0.494 & 0.729 \\%& \bf 0.375 & 0.153 & \bf 0.463 \\

\bottomrule

\bf t-BGER & \bf 0.428 & \bf 0.307 & \bf 0.444 & \bf 0.252 & \bf 0.142 & \bf 0.233 & \bf 0.741 & \bf 0.519 & \bf 0.998 & \bf 0.651 & \bf 0.500 & \bf 0.750 \\%& 0.324 & 0.083 & 0.375 \\

    \bottomrule
  \end{tabular}
\end{table*}

\subsection{Evaluation Metrics}
We considered three widely used rank evaluation metrics in literature to evaluate our approach. The ground truth is the user who provided the accepted answer for the question in the test dataset. The metrics include: \textbf{(i) Mean Reciprocal Rank (MRR)}: $= \frac{1}{Q}\sum_{q\in {Q}}\frac{1}{rank_q}$ - the average multiplicative inverse of the rank of the correct answerer, where $Q$ is the number of test samples and $rank_q$ is the rank of the answerer who provided the accepted answer (ground truth). \textbf{(ii) Precision@K}: The predicted instances where the ground truth answerer appeared in the top K rank list. We considered $K=1,3$ and reported P@1, P@3 - that is, where the answerer who provided the accepted answer appears in top 1 and top 3 ranked list respectively.

\subsection{Comparative methods}
We compare our method to the following state-of-the-art: collaborative \textbf{matrix-factorization} representative - (i) TAG-MF~\cite{yang2014tag} and deep-learning representatives - (ii) \textbf{network embedding based} NeRank~\cite{li2019personalized}, (iii) \textbf{transformer-based} method TCQR ~\cite{zhang2020temporal}. 
Furthermore, we will use two additional \textbf{baselines}: (iv) Score and (v) t-TAG-MF, defined by us to further promote explainability. In the following we define each of them:
(i) \textbf{TAG-MF} \cite{yang2014tag} In this approach user-tag activity matrix is decomposed to learn user and tag latent features and subsequently user expertise on a given tag. The number of latent features is set to 10, as was done in  previous studies~\cite{yang2014tag};
(ii) \textbf{NeRank} \cite{li2019personalized} This approach jointly learns features of questions contents, raisers and experts via a heterogeneous network embedding algorithm and utilizes Convolutional Neural Network to compute ranking scores; 
(iii) \textbf{TCQR} \cite{zhang2020temporal} The method utilizes a temporal context-aware model in multiple granularities of temporal dynamics to realize the multi-faceted and temporal-aware expert learning. 
The content of the question is encoded by a BERT~\cite{BERT} model (pre-trained deep bidirectional Transformers model);
(iv) \textbf{Score} This approach rank the users based on the number of positive votes minus the number of negative votes, averaged over all the answers provided by the user. Then we use the vote score as the probability to randomly generate the rank list of all the answerers and average the results by 50 trials.;
(v) \textbf{t-TAG-MF} Similar to TAG-MF approach, however the entries in the user-tag activity matrix are computed using temporal discounting. 

For temporal discounting the time window is set as one month in the equations ~\ref{eq_hyper_disc}, ~\ref{temporal_activity_Netscr}.

\subsection{Results}

Table \ref{perform} summarizes the results of our temporal weighted bipartite graph model on five real-world datasets. The results show that our model significantly outperforms all the baselines on all metrics. On average our model achieves a 12\% performance gain on MRR metrics compared to the best baseline method.
Further we repeat the experiment for our approach and t-NMF approach (the best performing baseline) with all the users, that is any user who has atleast one answer in the training set. Next we compare the performance of the cold-start users, that is the users with less than 10 answers in the training set (Table \ref{coldU}). We find that our approach outperforms the t-NMF achieving an average of 30\% performance gain on MRR metrics compared to t-NMF approach. This shows that for users with limited historical records our approach rank them significantly better.

\begin{table}
  \caption{Performance gain of t-BGER for cold start users compared to TAG-MF and t-TAG-MF on MRR metrics}
  \label{coldU}
  \begin{tabular}{|l|c|cl|cl|}
    \toprule

   & t-BGER & TAG-MF & gain & t-TAG-MF & gain\\
    \midrule

\bf philosophy & 0.043 & 0.003 & 14.1 & 0.033 & 1.3 \\
\bf history & 0.016 & 0.007 & 2.2 & 0.022 & 0.7 \\

\bf ebooks & 0.082 & 0.015 & 5.6 & 0.058 & 1.4 \\

\bf 3dprinting & 0.153 & 0.018 & 8.7 & 0.143 & 1.1 \\

%\bf ai & 0.057 & 0.010 & 5.9 & 0.028 & 2.0 \\

    \bottomrule
  \end{tabular}
  
\end{table}

\section{Conclusions}

In this paper, we propose a novel temporal weighted bipartite graph model for expert recommendation in CQA systems. Our model learns users' expertise in the context of both topic proficiency and temporal information to address the evolving interest and the activity patterns of users. Leveraging tags and bipartite graph approach our model performs better on highly sparse CQA systems where most users have limited historical records. This reduces the computational complexity of our approach making it easier for implementing periodic updates. Experiments on several real-world datasets demonstrated the advantages of our proposed model.

%%
%% The acknowledgments section is defined using the "acks" environment
%% (and NOT an unnumbered section). This ensures the proper
%% identification of the section in the article metadata, and the
%% consistent spelling of the heading.
%\begin{acks}
%To Robert, for the bagels and explaining CMYK %and color spaces.
%\end{acks}

%%
%% The next two lines define the bibliography style to be used, and
%% the bibliography file.
\bibliographystyle{unsrt}  
\bibliography{references}

\begin{thebibliography}{10}

\bibitem{srba2016comprehensive}
Ivan Srba and Maria Bielikova.
\newblock A comprehensive survey and classification of approaches for community
  question answering.
\newblock {\em ACM Transactions on the Web (TWEB)}, 10(3):1--63, 2016.

\bibitem{srba2016stack}
Ivan Srba and Maria Bielikova.
\newblock Why is stack overflow failing? preserving sustainability in community
  question answering.
\newblock {\em Ieee Software}, 33(4):80--89, 2016.

\bibitem{li2011question}
Baichuan Li, Irwin King, and Michael~R Lyu.
\newblock Question routing in community question answering: putting category in
  its place.
\newblock In {\em Proceedings of the 20th ACM international conference on
  Information and knowledge management}, pages 2041--2044, 2011.

\bibitem{neshati2017dynamicity}
Mahmood Neshati, Zohreh Fallahnejad, and Hamid Beigy.
\newblock On dynamicity of expert finding in community question answering.
\newblock {\em Information Processing \& Management}, 53(5):1026--1042, 2017.

\bibitem{zhou2012classification}
Tom~Chao Zhou, Michael~R Lyu, and Irwin King.
\newblock A classification-based approach to question routing in community
  question answering.
\newblock In {\em Proceedings of the 21st international conference on world
  wide web}, pages 783--790, 2012.

\bibitem{chang2013routing}
Shuo Chang and Aditya Pal.
\newblock Routing questions for collaborative answering in community question
  answering.
\newblock In {\em 2013 IEEE/ACM International Conference on Advances in Social
  Networks Analysis and Mining (ASONAM 2013)}, pages 494--501. IEEE, 2013.

\bibitem{zhang2007expertise}
Jun Zhang, Mark~S Ackerman, and Lada Adamic.
\newblock Expertise networks in online communities: structure and algorithms.
\newblock In {\em Proceedings of the 16th international conference on World
  Wide Web}, pages 221--230, 2007.

\bibitem{zhou2012topic}
Guangyou Zhou, Siwei Lai, Kang Liu, and Jun Zhao.
\newblock Topic-sensitive probabilistic model for expert finding in question
  answer communities.
\newblock In {\em Proceedings of the 21st ACM international conference on
  Information and knowledge management}, pages 1662--1666, 2012.

\bibitem{riahi2012finding}
Fatemeh Riahi, Zainab Zolaktaf, Mahdi Shafiei, and Evangelos Milios.
\newblock Finding expert users in community question answering.
\newblock In {\em Proceedings of the 21st international conference on world
  wide web}, pages 791--798, 2012.

\bibitem{li2015hybrid}
Hai Li, Songchang Jin, and LI~Shudong.
\newblock A hybrid model for experts finding in community question answering.
\newblock In {\em 2015 International Conference on Cyber-Enabled Distributed
  Computing and Knowledge Discovery}, pages 176--185. IEEE, 2015.

\bibitem{yang2013cqarank}
Liu Yang, Minghui Qiu, Swapna Gottipati, Feida Zhu, Jing Jiang, Huiping Sun,
  and Zhong Chen.
\newblock Cqarank: jointly model topics and expertise in community question
  answering.
\newblock In {\em Proceedings of the 22nd ACM international conference on
  Information \& Knowledge Management}, pages 99--108, 2013.

\bibitem{zhao2014expert}
Zhou Zhao, Lijun Zhang, Xiaofei He, and Wilfred Ng.
\newblock Expert finding for question answering via graph regularized matrix
  completion.
\newblock {\em IEEE Transactions on Knowledge and Data Engineering},
  27(4):993--1004, 2014.

\bibitem{wang2018survey}
Xianzhi Wang, Chaoran Huang, Lina Yao, Boualem Benatallah, and Manqing Dong.
\newblock A survey on expert recommendation in community question answering.
\newblock {\em Journal of Computer Science and Technology}, 33(4):625--653,
  2018.

\bibitem{pal2012evolution}
Aditya Pal, Shuo Chang, and Joseph Konstan.
\newblock Evolution of experts in question answering communities.
\newblock In {\em Proceedings of the International AAAI Conference on Web and
  Social Media}, volume~6, pages 274--281, 2012.

\bibitem{le2016retrieving}
Long~T Le and Chirag Shah.
\newblock Retrieving rising stars in focused community question-answering.
\newblock In {\em Asian Conference on Intelligent Information and Database
  Systems}, pages 25--36. Springer, 2016.

\bibitem{yang2014tag}
Baoguo Yang and Suresh Manandhar.
\newblock Tag-based expert recommendation in community question answering.
\newblock In {\em 2014 IEEE/ACM International Conference on Advances in Social
  Networks Analysis and Mining (ASONAM 2014)}, pages 960--963. IEEE, 2014.

\bibitem{zhou2010solving}
Tao Zhou, Zolt{\'a}n Kuscsik, Jian-Guo Liu, Mat{\'u}{\v{s}} Medo,
  Joseph~Rushton Wakeling, and Yi-Cheng Zhang.
\newblock Solving the apparent diversity-accuracy dilemma of recommender
  systems.
\newblock {\em Proceedings of the National Academy of Sciences},
  107(10):4511--4515, 2010.

\bibitem{zhou2007bipartite}
Tao Zhou, Jie Ren, Mat{\'u}{\v{s}} Medo, and Yi-Cheng Zhang.
\newblock Bipartite network projection and personal recommendation.
\newblock {\em Physical review E}, 76(4):046115, 2007.

\bibitem{green1994temporal}
Leonard Green, Nathanael Fristoe, and Joel Myerson.
\newblock Temporal discounting and preference reversals in choice between
  delayed outcomes.
\newblock {\em Psychonomic Bulletin \& Review}, 1(3):383--389, 1994.

\bibitem{ou2007power}
Qing Ou, Ying-Di Jin, Tao Zhou, Bing-Hong Wang, and Bao-Qun Yin.
\newblock Power-law strength-degree correlation from resource-allocation
  dynamics on weighted networks.
\newblock {\em Physical Review E}, 75(2):021102, 2007.

\bibitem{li2019personalized}
Zeyu Li, Jyun-Yu Jiang, Yizhou Sun, and Wei Wang.
\newblock Personalized question routing via heterogeneous network embedding.
\newblock In {\em Proceedings of the AAAI Conference on Artificial
  Intelligence}, volume~33, pages 192--199, 2019.

\bibitem{zhang2020temporal}
Xuchao Zhang, Wei Cheng, Bo~Zong, Yuncong Chen, Jianwu Xu, Ding Li, and Haifeng
  Chen.
\newblock Temporal context-aware representation learning for question routing.
\newblock In {\em Proceedings of the 13th International Conference on Web
  Search and Data Mining}, pages 753--761, 2020.

\bibitem{BERT}
Jacob Devlin, Ming-Wei Chang, Kenton Lee, and Kristina Toutanova.
\newblock Bert: Pre-training of deep bidirectional transformers for language
  understanding.
\newblock {\em arXiv preprint arXiv:1810.04805}, 2018.

\end{thebibliography}

%%
%% If your work has an appendix, this is the place to put it.
\appendix

\end{document}